# Surface Magnetism in $Fe_3GeTe_2$ Crystals

T. A. Tyson[1,5,*], S. Amarasinghe[1], AM M. Abeykoon[2], R. Lalancette[3], S. K. Du[4,5], X. Fang[4,5],

S.-W. Cheong[4,5], A. Al-Mahboob[6], and J. T. Sadowski[6,*]

[1]Department of Physics, New Jersey Institute of Technology, Newark, NJ  07102

[2]National Synchrotron Light Source II, Brookhaven National Laboratory, Upton, NY 11973

[3]Department of Chemistry, Rutgers University, Newark, New Jersey 07102

[4]Department of Physics and Astronomy, Rutgers University, Piscataway, NJ 08854

[5]Rutgers Center for Emergent Materials, Rutgers University, Piscataway, NJ 08854

[6]Center for Functional Nanomaterials, Brookhaven National Lab, Upton, New York 11973

*Corresponding Authors:

T. A. Tyson, e-mail: tyson@njit.edu

J. T. Sadowski, e-mail: sadowski@bnl.gov

**Abstract**

The surface magnetization of $Fe_3GeTe_2$ was examined by low-energy electron microscopy (LEEM) using an off-normal incidence electron beam. We found that the 180° domain walls are of Bloch type. Temperature-dependent LEEM measurements yield a surface magnetization with a surface critical exponent $\beta_1 = 0.79 \pm 0.02$. This result is consistent with surface magnetism in the 3D semi-infinite Heisenberg ($\beta_1 = 0.84 \pm 0.01$) or Ising ($\beta_1 = 0.78 \pm 0.02$) models, which is distinctly different from the bulk exponent ($\beta = 0.34 \pm 0.07$). The measurements reveal the power of LEEM with a tilted beam to determine magnetic domain structure in quantum materials. Single crystal diffraction measurements reveal inversion symmetry-breaking weak peaks and yield space group P-6m2. This Fe site defect-derived loss of inversion symmetry enables the formation of skyrmions in this $Fe_3GeTe_2$ crystal.

**Introduction**

Magnetism is critical to information storage and is one of the most commonly experienced physical properties. Despite this familiarity, the origin of this property is the result of very complex and fundamental phenomena resulting from the electron's intrinsic spin and interactions of electrons with each other. Intricate magnetic ordering of the magnetic moments generated by the atomic spin configurations yields ferromagnetic, antiferromagnetic, ferrimagnetic, vortex, chiral, and other diverse spin ordering patterns. Most recently, research efforts have focused on the utilization of nanoscale skyrmions and other complex swirling magnetic configurations to store and retrieve data in more energy-efficient and higher-density approaches than presently utilized [1,2,3,4].

From the fundamental physics perspective, well-established spin ordering models have been developed to predict the magnetic properties, including the Ising, XY, and Heisenberg models, which treat the localized spins as scalars, 2-dimensional vectors, or d-dimensional vectors, respectively [5,6,7 8]. The magnetic ordering of these spins in real space has been explored, and, in particular, the 1D and 2D Heisenberg models with isotropic spins S with finite-range interactions were predicted not to support ferromagnetic or antiferromagnetic orders at nonzero temperatures (no spontaneous magnetization) [9]. On the other hand, predictions based on the 2D XY (d=2 Heisenberg model) model reveal the presence of metastable topological long-range order related to the spin-spin correlations [10, 11]. This order manifests itself by forming vortex spin/antivortex spin pairs. Magnetic anisotropy opens up a gap in the spin wave spectrum, enabling this order at finite temperatures. As the first clear and recent example, the van der Waals system $CrCl_3$ grown as a monolayer on graphitized 6H-SiC(0001) was found by Cr L3-Edge x-ray magnetic circular dichroism (XMCD) to be an easy-plane 2D magnetic system with critical exponent $\beta = 0.227\pm0.021$ ($M = M_0[1-T/T_c]^\beta$ for $T < T_c$) [12], close to theoretical value of 0.231 for the 2D XY model [13].

The progress in fabrications of van der Waals systems and heterostructures composed of these materials has opened up the study of magnetism in reduced dimension by enabling the creation of materials with atomically abrupt and well-defined interfaces [14,15,16,17,18].These results were ushered in with the observation of 2D magnetism in the $Cr_2Ge_2Te_6$ and $CrI_3$ systems, which are 2D Heisenberg ferromagnets and 2D Ising antiferromagnets, respectively [19,20].These 2D van der Waals systems are of importance due to their ability to reveal the proper behavior of materials in reduced dimension but also for the possibility of new applications, including the ease of implementation of stimuli such as electric fields and strain, property tailoring by defect engineering, the ability to readily functionalize the surface and the possibility of making flexible materials systems [21,22,23,24].

Most magnetic van der Waals systems explored to date order magnetically at temperatures significantly below room temperature and must be modified for device applications. However, the $Fe_3GeTe_2$ system has been found to exhibit intriguing basic physics properties and yet have a magnetic transition above 200 K. This system is composed of four atom-thick van der Waals layers in which Fe is coordinated to Te. This material was first synthesized and described by structure (with assigned space centrosymmetric group $P6_3mmc$), electron transport, and magnetic properties ($T_c \sim 250$ K) in 2006 [25]. Detailed measurements on micrometer scale flakes with thicknesses down to one van der Waals layer reveal out-of-plane uniaxial magnetic anisotropy and a transition from 3D to 2D Ising behavior, with the transitions occurring near five der Waals layers of thickness [26]. The critical exponent $\beta$ was examined via magnetic circular dichroism measurements of magnetization, yielding $\beta = 0.14 \pm 0.02$ for the monolayer system and $\beta$ in the range 0.25 to 0.27 for thicker flakes. We note that the mean-field (independent of dimension), 2D Ising, 2D XY, and 2D Heisenberg models yield $\beta$ values of 0.5, 0.125, 0.23, and 0.5, while the corresponding 3D models have values of 0.5, 0.3265, 0.345, and 0.365, respectively [27,28,29]. For bulk samples, fits of the magnetization vs. field data at a set of fixed temperatures yielded $\beta = 0.327 \pm 0.003$, similar to the 3D Heisenberg model. However, the critical parameter $\gamma$ related to the magnetic susceptibility was found to be closest to that of a mean-field model [30]. Similar work suggests complete compatibility with the 3D Heisenberg model [31].The layer-dependent critical temperature was extracted from magnetization derived from Hall resistance measurements vs. magnetic field. Using these results, the spin-spin correlation range was estimated to be approximately five van der Waals layers [32].

Besides the space filing models (no surfaces) used to describe magnetic order, theoretical models that explicitly treat systems that have both surface and bulk regions (semi-infinite models) have been explored. In these models, near the magnetic ordering temperature, the bulk (interior) and surface magnetization take the form $M_B = M_{0B}[1-T/T_{c-B}]^{\beta}$ and $M_S = M_{0S}[1-T/T_{c-S}]^{\beta 1}$, respectively [33,34,35,36,37]. Distinct transition temperatures may occur for each region (bulk vs. surface) characterized by different exponents $\beta$, $\beta 1$ and critical temperatures, $T_{c-B}$ and $T_{c-S}$ of the bulk and surface regions. It is found that the semi-infinite 3D mean-field model has $\beta = 0.5$ with $\beta 1 = 1.0$ while the corresponding 3D Heisenberg and 3D Ising models have values of $\beta \approx 0.37$ with $\beta 1 \approx 0.84$ and $\beta = 0.3125$ with $\beta 1 \approx 0.78$, respectively.

The surface magnetization of thin films and crystals has been studied in simple metallic systems in the past in systems such as Gd metal, EuS(111) on Si(111), and Ni(110), mainly by spin-polarized electron scattering methods [34,38,39]. The EuS(111) system revealed that the bulk magnetic region (examined by magneto-optical Kerr measurements) and surface magnetic region (studied by spin-polarized LEED) order magnetically at the same temperature with surface region critical exponent $\beta 1 \approx 0.72(3)$.

In the present work, we first conducted single crystal diffraction measurements on $Fe_3GeTe_2$ and found weak peaks that violate the reflection conditions for the frequently quoted P63/mmm space group. The highest symmetry space group compatible with the observed single crystal data is the noncentrosymmetric space group P-6m2.

By utilizing low-energy electron microscopy (LEEM) with a tilted incident electron beam, we probed the surface to determine the magnetic domain boundaries, determine the type of domain walls (Neel vs. Bloch), and use the domain images (recorded in function of temperature) to extract the critical exponents $\beta 1$ for the surface magnetism in a $Fe_3GeTe_2$ crystal. The LEEM approach utilized here is called a tilted beam configuration [40], building on the development of LEEM by Bauer (see Ref. [41]). We extended this approach by using a simple model taking account of the Lorentz forces on electrons and the limited mean free path for electrons moving in real materials. This approach does not require the use of spin-polarized electrons. A comparison of low-temperature XMCD measurements with LEEM measurements with a tilted electron beam revealed that the same domain patterns are present in both images. A simple model of the interaction of the electron with the surface indicated that the domain walls are Bloch-type. Temperature-dependent LEEM measurements yielded a surface magnetization with a critical exponent $\beta_1 = 0.79 \pm 0.02$. This result is consistent with the 3D models for a semi-infinite magnetic system with surface magnetism $\beta_1 = 0.84 \pm 0.01$ (Heisenberg) or $\beta_1 = 0.78 \pm 0.02$ (Ising).

**Results**

**Crystal structure of $Fe_3GeTe_2$**

The crystal structure of the $Fe_3GeTe_2$ system was revisited to understand its properties fully. Six-fold redundant single crystal datasets (complete sphere measurements in reciprocal space) taken at 100 K and 275 K with Mo K$\alpha$ radiation revealed weak peaks, which violate the reflection condition for the currently used centrosymmetric $P6_3/mmc$ structure [26]. These weak peaks are difficult to observe in standard powder x-ray diffraction measurements unless long counting times and detectors with high dynamic range are utilized. The highest symmetry space group consistent with the observed reflections was found to be P-6m2 (Tables S1 and S2 of the Supplementary Information [44]). This noncentrosymmetric space group is a subgroup of $P6_3/mmc$.

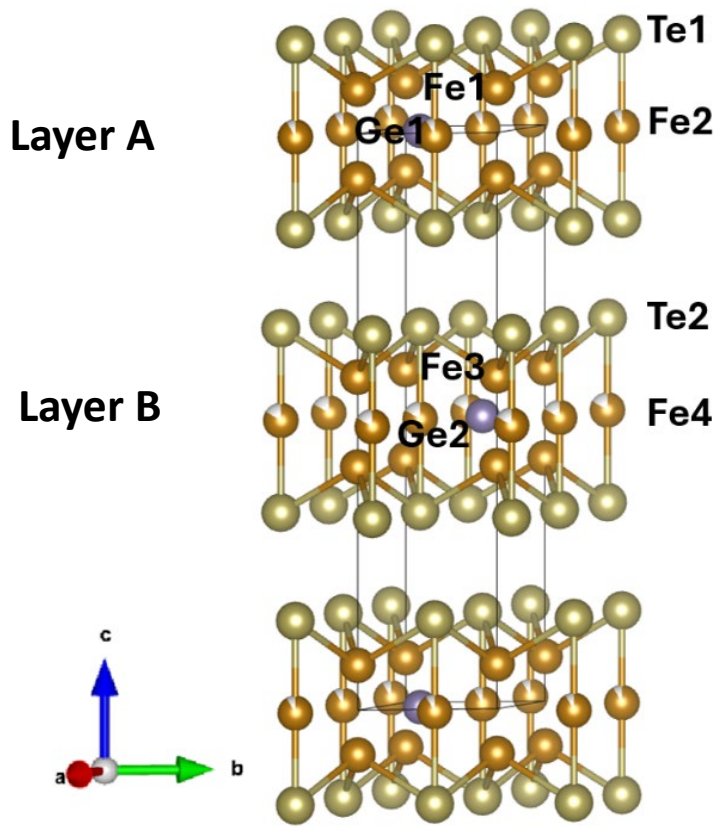

**Fig. 1.** Unit cell of "Fe$_3$TeGe$_2$" for the (non-centrosymmetric) space group P-6m2 space group solution showing the repeated AB stacking. The Fe2 site has an occupancy of 0.94 ± 0.03, and the Fe4 site has occupancy 0.80 ± 0.03. The actual sample composition is Fe$_{2.87}$TeGe$_2$.

In Fig. 1, we show the hexagonal unit cell with stacked layers of Te-coordinated Fe sites with Ge embedded in the layers. Unlike the P6$_3$/mmc space group, in the polar noncentrosymmetric P-6m2 space group, each unit cell has two unique layers (A and B in Fig. 1), with each layer containing distinct pairs of Fe sites Fe1/Fe2 and Fe3/Fe4 for layer A and B, respectively. The Fe2 site (layer A) has an occupancy of 0.94 ±0.03, while the Fe4 site (layer B) has an occupancy of 0.80 ±0.03. The real chemical composition of the crystal studied is thus Fe$_{2.87}$GeTe$_2$. Single crystal data acquisition and reduction are detailed in the supplementary document by the methods in Refs. [42,43,44,45,46,47]. It has to be noted that when x-ray powder diffraction measurement methods are applied to this material, the peaks that violate the P6$_3$/mmc space group symmetry can be seen in laboratory instruments only if long-time scans are carried out, or if measurements are conducted with a synchrotron source (see Fig. S2 in the Supplementary Information).

**NEXAFS and XMCD measurements**

To ascertain the chemistry of the surface we have performed near-edge x-ray absorption spectroscopy (NEXAFS) measurements in the XPEEM system. In Fig. 2a, we show the room temperature spectra of Fe L3 and L2 edges (~707 eV and ~720 eV) of the sample. It is evident that there is no strong oxygen-related peak [48] near the main Fe L3-Edge peak. Furthermore, we have performed low-temperature x-ray magnetic circular dichroism (XMCD) imaging of the surface magnetic structure. In Fig. 2b, we show the XMCD image of the same sample at 110 K (30 μm x 30 μm area) with bright and dark regions corresponding to magnetic moment parallel and antiparallel to the surface normal of the crystal, respectively.

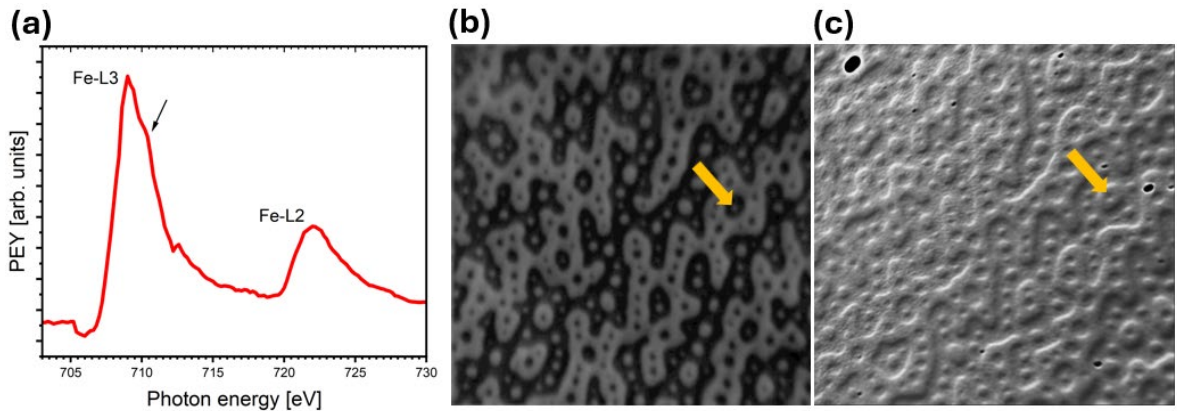

**Fig. 2.** (a) Room temperature right circularly polarized Fe L3/L2-edge spectrum of the $Fe_3GeTe_2$ crystal with an arrow pointing to weak main peak right shoulder, which is indicative of a sample with negligible oxide contamination. (b) XMCD image for 30 μm x 30 μm area taken at 110 K. The dark and bright regions correspond to up and down magnetization orientation relative to the sample normal. XMCD images were taken at the peak of the Fe L3-edge spectrum. (c) A tilted-beam LEEM image showing a surface magnetic contrast. The arrows in (b) and (c) point to the same domain feature. In the LEEM image, the arrow also represents the incident electron beam direction.

**XPEEM and LEEM characterization**

Systematic studies have previously been conducted to understand the nature of the magnetism in $Fe_3GeTe_2$ using anomalous Hall measurements, reflection magnetic circular dichroism, squid magnetometry, magnetic force microscopy, and XMCD measurements [15,27,32,34]. These measurements provide information on the relative magnetization amplitude vs. temperature. The spatial variation of the magnetic structure has been imaged by scanning tunneling microscopy, magnetic force microscopy, XMCD, scanning electron microscopy with polarization analysis imaging, and Lorentz transmission electron microscopy [27,49,50,51,52,53]. The domain pattern in the absence of an external magnetic field follows the predicted pattern observed for thick crystals with uniaxial magnetic anisotropy [54,55,56].

Unlike the techniques previously applied to these materials, LEEM measurements with a tilted incident electron beam do not require special sample preparation or a spin-polarized electron source as in the case of spin-polarized LEEM experiments [57,58,59,60]. Magnetic contrast can be observed by physically tilting the incident electron beam relative to the sample surface normal (or tilting the sample). We show here how this pattern can be interpreted to extract the domain wall type and determine critical exponents of the surface magnetism. The resulting measurement produces the contrast seen in Fig. 2c for the same sample region. The sample probing depth in LEEM is ≈ 1 nm compared to ≈ 10 nm for XMCD, resulting in the measurement of the surface magnetization contrast. In Fig. 2b and Fig. 2c, the arrows point to the same domain feature shifted in figures by a thermal drift of the sample.

In addition, in the LEEM image, the arrow also represents the incident electron beam direction. Comparing the XMCD image in Fig. 2b with the LEEM image in Fig. 2c, we see very close matching of the magnetic domain contrast. A cartoon of the domain structure with wavy patterns for crystals above the critical thickness of Kittel-type structure (left) compared to the pattern for thickness beyond this critical length (right) is shown in Fig. 3a. This domain pattern is well understood, but the domain wall nature and the critical exponents of surface magnetism are not well known. Here, we will apply a simple model to understand the domain wall structure.

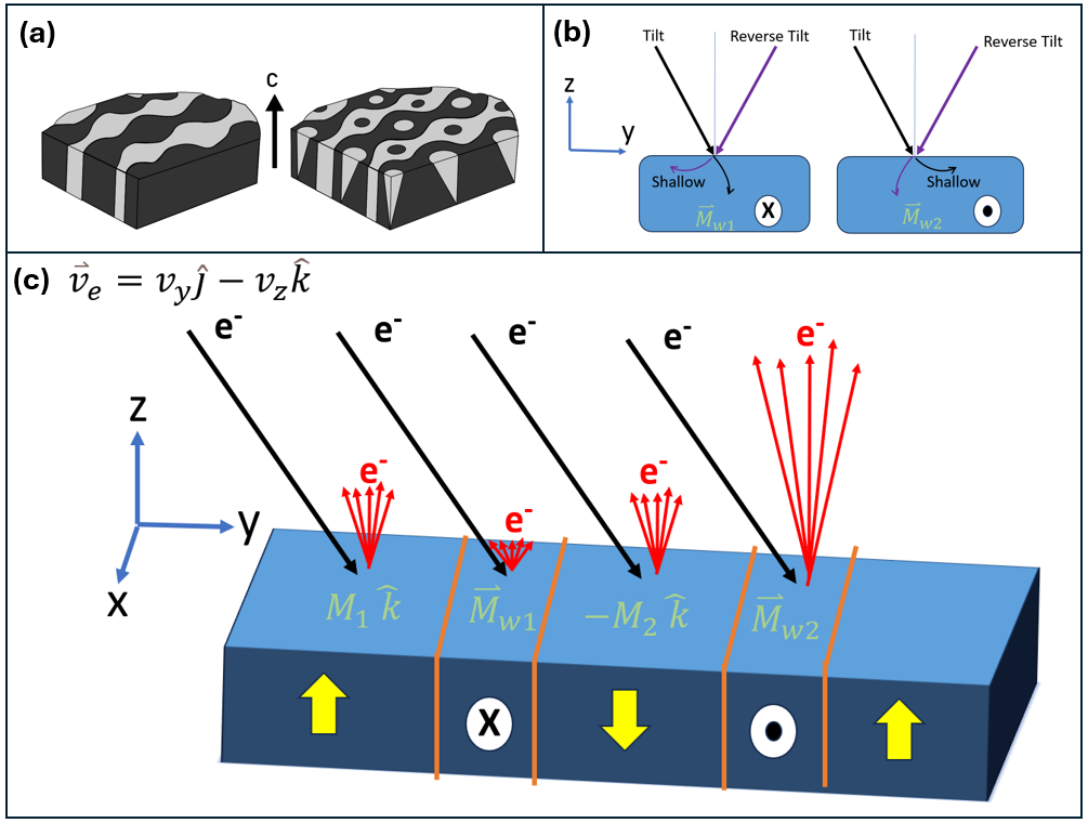

**Fig. 3.** (a) A sketch of a magnetic domain structure in $Fe_3GeTe_2$ with wavy patterns for crystals above the critical thickness of Kittel-type structure (left) compared to pattern for thickness beyond this critical length (right). (b) A sketch illustrating that by considering the Lorentz force region within the domain wall (Bloch type), the deflection of the beam relative to the sample normal can be enhanced or reduced, depending on the tilt of the electron beam. (c) Schematic of LEEM measurement with a tilted beam showing the expected electron return flux at the detector (red lines) for distinct domain and domain wall regions for given incident electron beam directions (black arrows and purple arrows). From left to right: a section of the sample with a domain with magnetization along z ($M_1 \hat{k}$), which switches orientation via a Bloch domain wall ($\vec{M}_{w1}$), to a domain magnetized along -z (-$M_2 \hat{k}$)) which then reverses direction via a Bloch domain wall ($\vec{M}_{w2}$). Electrons in the incident beam have directions defined by $\vec{v}_e = v_y \hat{j} - v_z \hat{k}$.

Fig. 3b shows that by determining the Lorentz force direction in each region, the deflection of the electron beam relative to the sample normal (and magnetic contrast) can be ascertained. For the first domain wall ($\vec{M}_{w1}$), the electrons incident on the sample will penetrate more deeply (deflected toward the sample normal) while in the second domain wall ($\vec{M}_{w2}$), electrons will suffer enhanced deflection from the normal sample. In the absence of atoms (i.e., in a vacuum), the electrons will spiral about the local magnetic field direction. However, in the presence of scattering centers (defects) and channels for energy loss, electrons that enter deeply into the materials will be heavily attenuated in intensity on returning to the surface. Electrons that spiral toward the surface suffer significantly less attenuation. This will produce an enhanced signal at the detector above the sample plane. Deeper penetration will result in signal suppression at the detector.

A schematic of LEEM measurements with a tilted beam showing the expected electron return flux at the detector (red lines) for distinct domain and domain wall regions for given incident electron beam directions (black arrows and purple arrows) is shown if Fig. 3c. Using the Lorentz force ($\vec{F}_e = q\left(\vec{v_e} \times \vec{B}\right) = q\,\mu_0\left(\vec{v_e} \times \vec{M}\right)$) on an electron with velocity $\vec{v_e}$ due to the local magnetic field $\vec{B}$ which is provided by the magnetization ($\vec{M}$) of the material, we can develop a simple model for the observed pattern. From left to right (in Fig. 3a), we show a section of the sample with a domain with magnetization along z ($M_1\,\hat{k}$), which switches orientation via a Bloch domain wall ($\vec{M}_{w1}$), to a domain magnetized along -z ($-M_2\,\hat{k}$)) which then reverses direction via a Bloch domain wall ($\vec{M}_{w2}$). Electrons in the incident beam have directions defined by $\vec{v}_e = v_y\hat{j} - v_z\hat{k}$. Reversing the beam direction ($\vec{v}_e = -v_y\hat{j} - v_z\hat{k}$) will reverse the domain contrast. No contrast difference will be seen between the domains with magnetization, $M_1\,\hat{k}$ and $-M_2\,\hat{k}$. However, they should produce a signal at the detector significantly higher than the deep penetration electrons in the domain wall region. Hence, the contrast seen in the images in Fig. 2c is at the domain wall boundary, with bright and dark features defining these boundaries. We note that in the model, we utilized a Bloch-type domain wall model with the magnetization vector rotating parallel to the domain wall. The use of a Neel-type wall (magnetization vector rotating normal to the domain wall) will produce no contrast.

To further support our model, we show LEEM images for a ≈ 13 μm x 13 μm expanded area of the sample with a tilted incident electron beam in Fig. 4a and for the reverse tilt configuration in Fig. 4b, respectively. As predicted from our model, reversing the tilt angle results in a reversal of the intensity pattern in the domain walls. To quantify the observed contrast, the line scans in Fig. 4c show the relative intensities in the domains for both beam tilt directions. The LEEM images obtained under normal incidence do not show any contrast, neither in the domain walls nor in the domains themselves (see Fig. S3 in the Supplementary Materials).

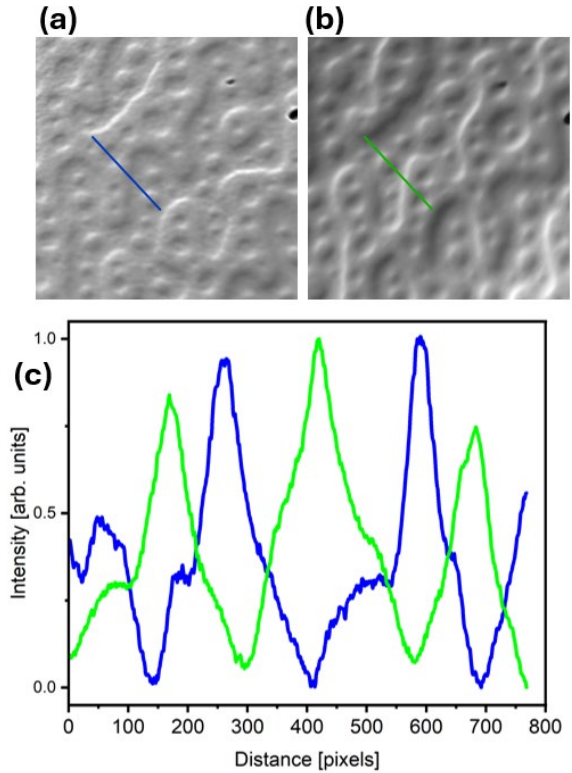

**Fig. 4.** LEEM measurements for a ≈ 13 μm x 13 μm expanded area of the sample with (a) tilt and (b) reverse tilt configurations showing the reversal of intensity in the domain wall regions as predicted in Fig. 3. (c) Line scans showing the relative intensities along the lines in the domains for tilted-beam (blue) and reversed tilted-beam (green) directions.

**Surface Magnetization**

As demonstrated in the model above, the tilted-beam LEEM contrast is derived from the sample surface. We have also established that the domain near the surface are of Bloch-Type (and possibly this exists in the interior bulk region). We will now utilize the tilted-beam LEEM images to extract the temperature dependence of the surface magnetization and surface critical exponents (β1).

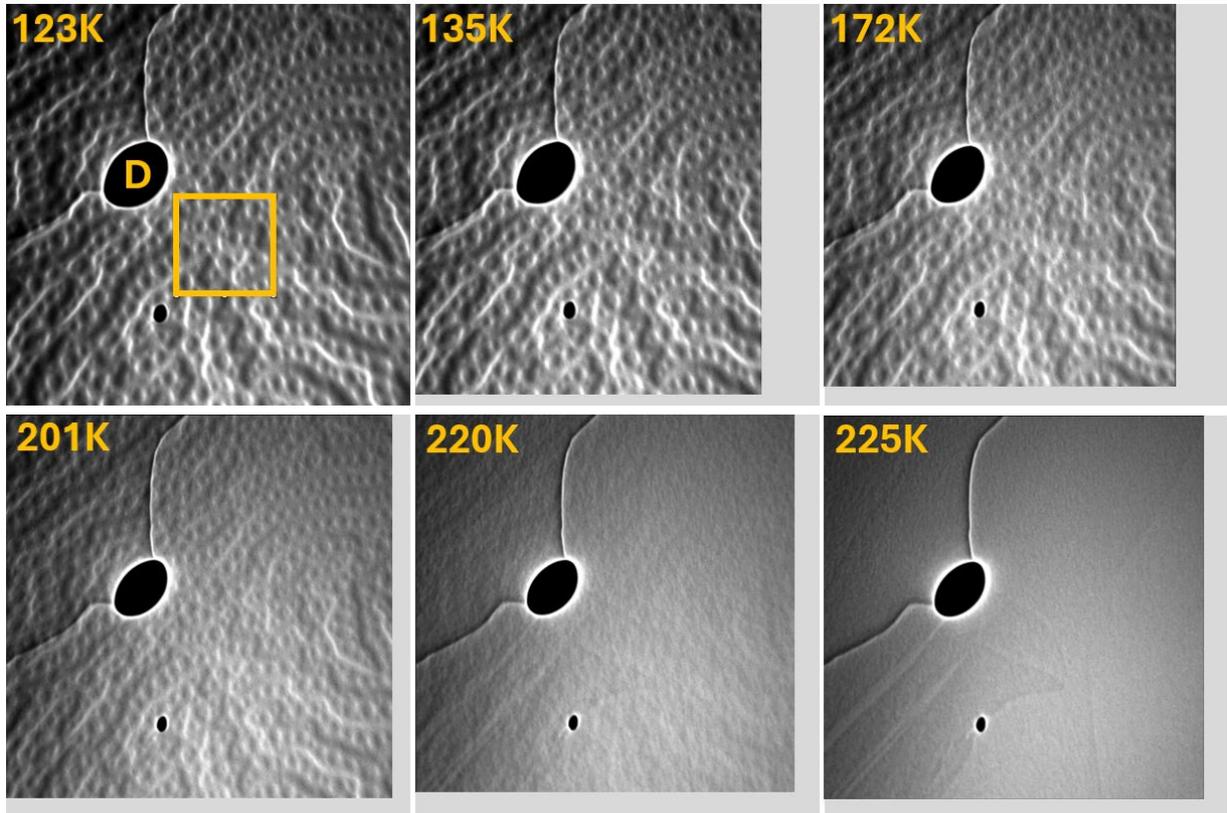

**Fig. 5.** Temperature-dependent tilted-beam LEEM images showing the surface magnetic domain structure with defect marker (D) used for a thermal drift correction alignment between the images. A fixed region of interest (ROI), defined by the yellow box, was used for each image to determine the temperature-dependent pixel amplitude standard deviation shown in Fig. 6. This standard deviation (StdDev) for each image was used as a proxy for surface magnetization. The ROI box has dimensions of 12μm x 12μm. The full square image area in each panel is 50μm x 50μm.

Fig. 5 gives a subset of the measured temperature-dependent tilted-beam LEEM images showing the surface magnetic domain structure with a defect marker (D) serving as a reference for a thermal drift correction between the images. A fixed region of interest (ROI) defined by the yellow box with dimensions of 12 μm x 12 μm, was used for each image to determine the pixel amplitude standard deviation for each image. This standard deviation (StdDev) at each pixel (relative to the average pixel intensity) in each image (one image per temperature) is as a proxy for surface magnetization.

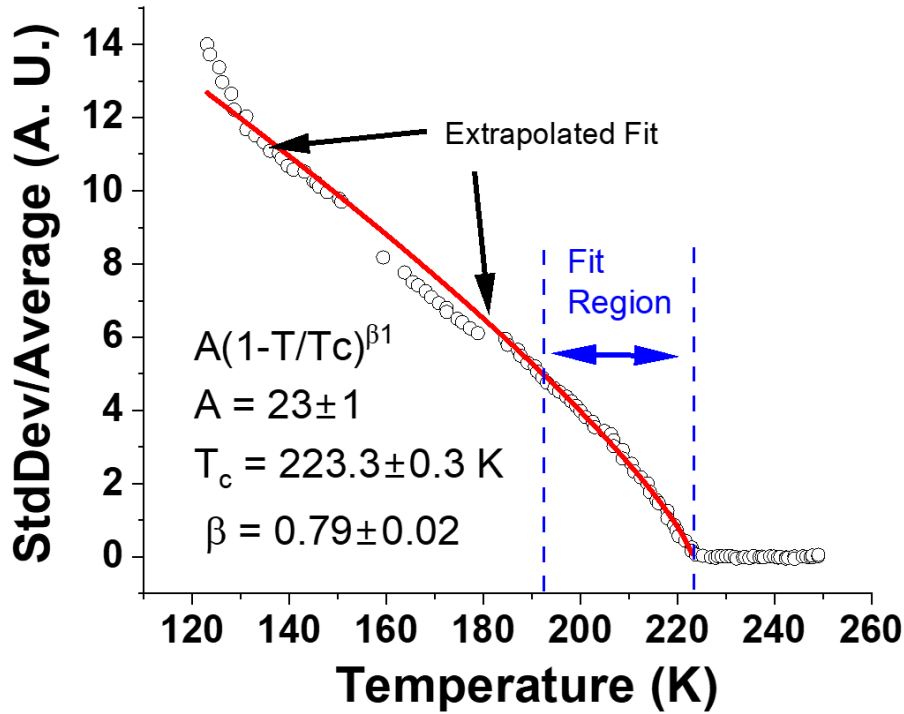

**Fig. 6.** Power law fit $(A*(1-T/T_c)^{b1})$ to the ROI pixel standard deviation (magnetic contrast, circles) vs. temperature (red line) yielding b1 = 0.79±0.02 and $T_c$ =223.2±0.03 K. Background above 225 K was subtracted. The fit over the region between 192 and 222 K is extrapolated to cover the full data range. Each data point (circle) corresponds to a full LEEM image, as in Fig 5.

The circles in Fig. 6 represent the ROI StdDev values for images taken at fixed temperatures. A power law fit (to $A*(1-T/Tc)^{\beta 1}$) over the region between 192 and 222 K is extrapolated to cover the full data range (red curve). The fit yielded $\beta 1$ = 0.79±0.02, and $T_c$ =223.2±0.03 K. Background above 225 K was subtracted. Each data point (circle) corresponds to the ROI of a full LEEM image, as in Fig 5. Bulk magnetization with data taken from SQUID measurements of Li *et al*. [69] is given in Fig. S4. For the bulk data, the fit was conducted between 170 K and 231 K and extrapolated to lower temperatures. The bulk fit yielded $\beta$ = 0.34±0.07 and $T_c$ =231±1 K.
14

**Discussion**

The crystal structure of the Fe$_3$GeTe$_2$ system was revisited, and a complete structural solution with a highly redundant set of reflections (six-fold redundant on complete reciprocal sphere data sets) has been obtained. The data were used to systematically explore the hexagonal space groups. The P-6m2 space group was found to be the highest symmetry space group, which gives no reflection absence violations. Structural analysis and the complete structures at 275 K and 100 K are given in the Supplementary Information. The Fe2 site (layer A) has an occupancy of 0.94 ±0.03, while the Fe4 site (layer B) has an occupancy of 0.80 ±0.03. Hence, within the experimental uncertainties, we can think of the A layer as essentially defect-free while the B layer has Fe vacancies (Fe4 sites).

We note that the P-6m2 space group is a subgroup of the currently used centrosymmetric P6$_3$/mmc space group. As shown in Fig. 1, two layers A and B with different Fe site occupancy exist. This breaks the inversion symmetry. We note that in the presence of a magnetic field, this system exhibits a skyrmion magnetic lattice structure [61,62,63] characterized by magnetic units (vortices) with spin orientation rotating continuously from up at the core, for example, to down away from the core as one moves radially out from the core. This new space group is compatible with the previously established space group but includes both strong and weak reflections (omitting the latter gives P6$_3$/mmc). It is also compatible with the observation of skyrmions in this material since the pioneering work of Bogdanov *et al*. [64] indicates that noncentrosymmetry is a necessary condition for forming a lattice of these types of magnetic vortices. The most stable vortices have varying magnetization magnitude, with a maximum occurring at the center. We now consider the surface magnetization hosted by this structure.

To date, techniques that can probe specifically the surface magnetism have not been applied to these van der Waals systems. Electron-based methods are ideal for examining surface magnetization. We note that in the absence of phonon scattering, the carrier mean free path ($\lambda$) in a material in terms of the Fermi wavevector ($k_F = (3\pi^2 n)^{1/3}$), resistivity ($\rho$), and carrier concentration (n) is given by $\lambda = \frac{3\hbar\pi^2}{\rho e^2 k_F^2}$ [65]. Observing that the resistivity of bulk Fe$_3$GeTe$_2$ is reported to be approximately 0.2 to 0.4 mΩ·cm (very poor metal) with weak temperature variation [66] and that the system has a Hall measurement derived carrier concentration of $\approx$ 3.3 x $10^{20}$ carriers/cm³, we estimate the carrier mean free path in these materials to be below 10 nm. We also note that the electron inelastic mean free path computed from the dielectric loss function yields a mean free path in simple metals and semiconductors of approximately 10 nm [67,68]. On the other hand, the penetration depth for photons at the Fe L3 edge in the case of x-ray magnetic circular dichroism (XMCD) measurements is approximately 100 nm. However, measurement of the electrons generated by the de-excitation process inside the sample is limited, as in the case of low energy electron microscopy (LEEM) by the electron mean free path. This result is consistent with the reported sampling depth of $\approx$ 10 nm for Fe L3 XMCD measurements on this sample [69]. The similarity between bulk magnetization and XMCD results can be seen in the beta exponents extracted when squid magnetization and XMCD measurements are combined.

The domain walls are seen in the surface contrast accessed by the LEEM measurements. As our model of tilted incident electron beam LEEM contrast revealed (Fig. 3), the domain walls on the surface are Bloch-type. This observation is consistent with the results from Ref. [70], where measurements using spin-polarized scanning tunneling microscopy (examining differential conductance) were modeled. The data in these measurements are consistent with Bloch-type wall of width $\approx$ 7 nm. This domain wall width can be extracted from the anisotropy constant $K_1$ and the exchange constant A as $d = \pi \left(\frac{A}{K_1}\right)^{1/2}$. Estimating the exchange constant by $A = 3 k_B T / na$ [50, 52] with n = 4 (coordination of magnetic atoms), a $\approx$ 2.6 Å (approximate Fe-Fe distance), and Tc =223 K and using an experimentally determined exchange constant $K_1 = 0.15$ x $10^7$ J/m³ [50, 52], we find the identical result of 7 nm for the domain wall boundary width. The similarity in the measured domain wall boundary width and this calculation indicate that the surface exchange constants are equivalent to the bulk values. The critical exponent of the surface β1 is, however, different. The similarity of the surface and bulk transition temperature indicated the deep extent of the magnetic domains at the surface into the sample interior region.

Surface magnetism in the Fe$_3$GeTe$_2$ system is distinctly different from the corresponding quantity in thin metallic films and nanoparticles. In those latter systems, the coordination of the magnetic ion by neighboring similar ions is reduced near the surface. In the case of the van der Waals system, each layer remains intact, even at the top layer. This explains why the transition temperature at the surface does not change (recall $A = 3\, k_B T/na$ is the same in each layer). Temperature-dependent tilted-beam LEEM measurements yield a surface magnetization with a surface critical exponent β1 = 0.79±0.02. This result is consistent with surface magnetism in the 3D semi-infinite Heisenberg (β1 = 0.84±0.01) or Ising (β1 = 0.78±0.02) models. This is distinctly different from the bulk exponent (β = 0.34±0.07). As noted above, for bulk samples, fits of the magnetization vs. field data at a set of fixed temperatures yielded β = 0.327 ±0.003, similar to the 3D Heisenberg model. The results thus indicate that in the absence of a magnetic field, the Heisenberg-type model of spin interactions is appropriate both at the surface and bulk of these materials.

**Conclusions**

The surface magnetization of Fe$_3$GeTe$_2$ was examined by low-energy electron microscopy (LEEM) using an off-normal incidence electron beam. We found that the 180º domain walls are of Bloch type. Temperature-dependent LEEM measurements yield a surface magnetization with a surface critical exponent β1 = 0.79±0.02. This result is consistent with surface magnetism in the 3D semi-infinite Heisenberg (β1 = 0.84±0.01) or Ising (β1 = 0.78±0.02) models. This is distinctly different from the bulk exponent β = 0.34±0.07). The measurements show the utility of LEEM measurement with a tilted beam in determining magnetic domain structure in quantum materials. Single crystal diffraction measurements exhibit weak inversion symmetry-breaking peaks, leading to a noncentrosymmetric space group P-6m2 in this material.

# Methods

**Sample Preparation and Conditions**

Single crystals of $Fe_3GeTe_2$ were prepared by chemical vapor transport. A small crystal (220 μm x 130 μm x 60 μm) from the main crystal was cleaved for use in single crystal diffraction measurements. For the LEEM and PEEM experiments, samples were cleaved in air by the scotch tape method and transferred to the vacuum system immediately (~1 minute) after treatment. The antechamber for sample transfer was at a pressure of $\approx 2 \times 10^{-9}$ Torr and the pressure of the UHV main chamber used in the measurements was $\approx 5 \times 10^{-10}$ Torr.

**Low Energy Electron Microscopy (LEEM) and X-ray Photoelectron Emission Microscopy (XPEEM)**

X-ray absorption spectroscopy (XAS) and LEEM measurements have been performed at the XPEEM/LEEM end station of the Electron Spectro-Microscopy beamline (ESM, 21-ID) at the National Synchrotron Light Source II (NSLS-II). All measurements were conducted using the aberration-corrected ELMITEC LEEM III system. XAS measurements were performed in a partial yield mode collecting the secondary electrons with the 2eV energy analyzer slit centered over the maximum of the secondary electron emission peak. In both, LEEM and PEEM measurements, the extractor field between the sample and objective lens was 20 keV. Pixel-wise XAS were obtained by recording a series of XPEEM images at each energy in a given absorption edge range at sequential increments of 0.2 eV, with the incident X-ray beam at ~17° to surface flat. The LEEM measurements were conducted in mirror electron microscopy mode with the incident electrons having low kinetic energy (~1 eV).

**Single Crystal Diffraction Measurements**

Single crystal data sets were collected at 275K and 100K on a 220 μm x 130 μm x 60 μm single crystal using a Rigaku XtaLAB Synergy-S X-ray diffractometer equipped with a HyPix-6000HE hybrid photon counting (HPC) detector and microfocused Mo-Kα radiation ($\lambda = 0.71073$ Å). The HPC detector has a high dynamic range with maximum count rate = $10^6$ cps/pixel (counter depth = 31 bit, 100 μm x 100 μm pixels, Si pixels each with CMOS-based readout electronics, direct photon counting, zero dark or readout noise). Use of Mo-Kα radiation resulted in a sampling depth of approximately 40 μm (full mass density). This should be compared to a sample dept of 6 μm (full density) for Cu-Kα radiation ($\lambda = 1.54184$ Å). The acquisition and reduction of the sixfold redundant datasets was conducted using CrysAlis[Pro] [44]. Scaling (SCALE3 ABSPACK scaling algorithm [45]) and analytical absorption corrections [46] were applied to the data (based on face-indexing in CrysAlis[Pro]). The structures were solved *via* intrinsic phasing methods using ShelXT and refined using ShelXL in the Olex2 graphical user interface.

**Powder Diffraction Data**

Powder diffraction data were collected using Beamline 28-ID-1 (PDF) at the NSLS-II, with a wavelength of 0.1665 Å in a 1 mm Kapton capillary sealed with clay. This bulk measurement can be compared with laboratory powder measurements using Cu Kα and Mo Kα radiation. These latter measurements only probe the surface of mm scale samples in the holders, with penetration depths near 6 μm and 40 μm, respectively (https://henke.lbl.gov/optical_constants/atten2.html). Note that Mo Kα radiation penetration is compatible with the single crystal used in the measurement above.

## Acknowledgments

This work is supported by NSF Grant No. DMR-1809931. Work at Rutgers University is supported by DOE Grant No. DE-GF02-07ER46382. The single crystal x-ray diffractometer was acquired under NSF MRI CHE-2018753 (Rutgers University, Newark). This research used resources of the Center for Functional Nanomaterials and the National Synchrotron Light Source II, which are U.S. Department of Energy (DOE) Office of Science facilities at Brookhaven National Laboratory, under Contract No. DE-SC0012704.

## Data availability

The data that support the findings of this study are available from the corresponding authors upon reasonable request.

## Authors contributions

T.A.T. and J.T.S. planned and supervised the research. S.K.D., X.F. and S.-W.C. conducted synthesis of $Fe_3GeTe_2$ single crystal. T.A.T. and R.L. conducted single crystal x-ray diffracton data collection and analysis. T.A.T., S.A. and AM M. A. conducted x-ray powder diffraction measurements and data analysis. T.A.T., A.A.M and J.T.S. performed LEEM and XAS measurements and data analysis. T.A.T. and J.T.S. prepared the manuscript. All the authors discussed the results and provided comments on the manuscript.

# References

[1] B. Göbel, I. Mertig and O. A. Tretiakov, "Beyond skyrmions: Review and perspectives of alternative magnetic quasiparticles", Physics Reports **895**, 1 (2021).

[2] Y. Tokura and N. Kanazawa, "Magnetic Skyrmion Materials", Chem. Rev. **121** (5), 2857 (2021).

[3] M. Hoffmann, G. P. Müller, C. Melcher and S. Blügel, "Skyrmion-Antiskyrmion Racetrack Memory in Rank-One DMI Materials", Front. Phys. **9** (2021).

[4] K. Wang, V. Bheemarasetty, J. Duan, S. Zhou and G. Xiao, "Fundamental physics and applications of skyrmions: A review", J Magn Magn Mater **563,** 169905 (2022).

[5] Stephen Blundell, *Magnetism in Condensed Matter*, (Oxford Press, New York, 2001).

[6] S. Friedli and Y. Velenik, *Statistical Mechanics of Lattice Systems: A Concrete Mathematical Introduction*, (Cambridge University Press, Cambridge, 2017).

[7] J. J. Binney, N. J. Dowrick, A. J. Fisher, and M. E. Newman, *The Theory of Critical Phenomena: An Introduction to the Renormalization Group*, (Clarendon Press, Cambridge, 1992).

[8] I. Herbut, A Modern Approach to Critical Phenomena, (Cambridge University Press, Cambridge, 2007).

[9] N. D. Mermin and H. Wagner, "Absence of ferromagnetism or antiferromagnetism in one- or two-dimensional isotropic Heisenberg models", Phys. Rev. Lett **17**, 1133 (1966).

[10] V. Berezinskiĭ, "Destruction of long-range order in one-dimensional and two-dimensional systems possessing a continuous symmetry group. II. Quantum systems" Sov Phys JETP **34**, 610 (1972)

[11] J. M. Kosterlitz and D. J. Thouless, "Ordering, metastability and phase transitions in two-dimensional systems",  J Phys C **6**, 1181 (1973)

[12] A. Bedoya-Pinto1, J.-R. Ji, A. K. Pandeya, P. Gargiani, M.Valvidares, P. Sessi, J. M. Taylor, F. Radu, K. Chang, S. S. P. Parkin, , "Intrinsic 2D-XY ferromagnetism in a van der Waals monolayer", Science **374**, 616 (2021).

[13] S. T. Bramwell and P. C. W. Holdsworth, "Magnetization and universal sub-critical behaviour in two-dimensional XY magnets", J. Phys. Condens. Matter **5**, L53 (1993).

[14] N. Samarth, "Magnetism in Flatland", Nature **546**, 216 (2017).

[15] Y. Li, B. Yang, S. Xu, B. Huang and W. Duan, "Emergent Phenomena in Magnetic Two-Dimensional Materials and van der Waals Heterostructures", ACS Applied Electronic Materials **4** (7), 3278 (2022).

[16] X. Song, F. Yuan and L. M. Schoop, "The properties and prospects of chemically exfoliated nanosheets for quantum materials in two dimensions", Applied Physics Reviews **8** (1), 0113121 (2021).

[17] X. Jiang, Q. Liu, J. Xing, N. Liu, Y. Guo, Z. Liu and J. Zhao, "Recent progress on 2D magnets: Fundamental mechanism, structural design and modification", Applied Physics Reviews **8** (3), 0313051 (2021).

[18] M. C. Wang, C. C. Huang, C. H. Cheung, C. Y. Chen, S. G. Tan, T. W. Huang, Y. Zhao, Y. Zhao, G. Wu, Y. P. Feng, H. C. Wu and C. R. Chang, "Prospects and Opportunities of 2D van der Waals Magnetic Systems", Annalen der Physik **532**, 1900452 (2020).

[19] C. Gong, L. Li, Z. Li, H. Ji, A. Stern, Y. Xia, T. Cao, W. Bao, C. Wang, Y. Wang, Z. Q. Qiu, R. J. Cava, S. G. Louie, J. Xia and X. Zhang, "Discovery of intrinsic ferromagnetism in two-dimensional van der Waals crystals", Nature **546**, 265 (2017).

[20] B. Huang, G. Clark, E. Navarro-Moratalla, D. R. Klein, R. Cheng, K. L. Seyler, D. Zhong, E. Schmidgall, M. A. McGuire, D. H. Cobden, W. Yao, D. Xiao, P. Jarillo-Herrero and X. Xu, "Layer-dependent ferromagnetism in a van der Waals crystal down to the monolayer limit", Nature **546** (7657), 270 (2017).

[21] C. Gong and X. Zhang, "Two-dimensional magnetic crystals and emergent heterostructure devices", Science **363** 6428 (2019).

[22] D. L. Cortie, G. L. Causer, K. C. Rule, H. Fritzsche, W. Kreuzpaintner and F. Klose, "Two-Dimensional Magnets: Forgotten History and Recent Progress towards Spintronic Applications", Advanced Functional Materials **30** (18), 1901414 (2019).

[23] S. J. Liang, B. Cheng, X. Cui and F. Miao, "Van der Waals Heterostructures for High-Performance Device Applications: Challenges and Opportunities", Adv Mater **32** (27), e1903800 (2020).

[24] M. C. Wang, C. C. Huang, C. H. Cheung, C. Y. Chen, S. G. Tan, T. W. Huang, Y. Zhao, Y. Zhao, G. Wu, Y. P. Feng, H. C. Wu and C. R. Chang, "Prospects and Opportunities of 2D van der Waals Magnetic Systems", Annalen der Physik **532** (5), 1900452 (2020).

[25] H. J. Deiseroth, K. Aleksandrov, C. Reiner, L. Kienle, R. K. Kremer, "$Fe_3GeTe_2$ and $Ni_3GeTe_2$ – Two New Layered Transition-Metal Compounds: Crystal Structures, HRTEM Investigations, and Magnetic and Electrical Properties", Eur. J. Inorg. Chem. 1561 (2006).

[26] Z. 26, B. Huang, P. Malinowski, W. Wang, T. Song, J. Sanchez, W. Yao, D. Xiao, X. Zhu, A. F. May, W. Wu, D. H. Cobden, J. H. Chu and X. Xu, "Two-dimensional itinerant ferromagnetism in atomically thin Fe(3)GeTe(2)", Nat Mater **17** (9), 778 (2018).

[27] A. Oleaga, A. Salazar, M. Ciomaga Hantean and G. Balakrishnan, "Three-dimensional Ising critical behavior in $R_{0.6}Sr_{0.4}MnO_3$(R=Pr,Nd)manganites", Physical Review B **92** (2), 0244091 (2015).

[28] D. Kim, B. L. Zink, F. Hellman and J. M. D. Coey, "Critical behavior of La0.75Sr0.25MnO3", Physical Review B **65** (21) (2002).

[29] A. Taroni, S. T. Bramwell and P. C. Holdsworth, "Universal window for two-dimensional critical exponents", J Phys Condens Matter 20 (27), 275233 (2008).

[30] B. Liu, Y. Zou, S. Zhou, L. Zhang, Z. Wang, H. Li, Z. Qu and Y. Zhang, "Critical behavior of the van der Waals bonded high T (C) ferromagnet Fe(3)GeTe(2)", Sci Rep **7** (1), 6184 (2017).

[31] Y. Liu, V. N. Ivanovski and C. Petrovic, "Critical behavior of the van der Waals bonded ferromagnet Fe$_{3-x}$GeTe$_2$", Physical Review B 96 (14), 1444291 (2017).

[32] C. Tan, J. Lee, S. G. Jung, T. Park, S. Albarakati, J. Partridge, M. R. Field, D. G. McCulloch, L. Wang and C. Lee, "Hard magnetic properties in nanoflake van der Waals Fe(3)GeTe(2)", Nat Commun **9** (1), 1554 (2018).

[33] P. J. Jensen and K. M. Bennemann, "Surface and Thin Film Magnetization of Transition Metals", Langmuir **12**, 45 (1996).

[34] K. Binder and D. P. Landau, "Crossover Scaling and Critical Behavior at the "Surface-Bulk" Multicritical Point", Phys. Rev. Lett. **52**, 318 (1984).

[35] H. W. Deihl and A. Nusser, "Critical Behavior of the Nonlinear σ. Model with a Free Surface: The "Ordinary" Transition in 2+ ε Dimensions", Phys. Rev. Lett. **56**, 2834 (1986).

[36] B. H. Dauth, S. F. Alvarado, M. Campagna, Critical Behavior of the Surface Magnetization of an Isotropic Heisenberg Ferromagnet: EuS(111) on Si(111)", Phys. Rev. Lett. **58**, 2118 (1987).

[37] C. S. Arnold and D. D. Pappas, "Gd(0001): Semiinfinite 3D Heisenberg Ferromagnet with Ordinary Surface", Phys. Rev. Lett. **85**, 5202 (2000).

[38] R. J. Celotta and R. T. Pierce, Polarized Electron Probes of Magnetic Surfaces", Science 234, 333 (1986).

[39] R. J. Celotta, D. T. Pierce, G.-C.Wang, S. D. Bader, and G. P. Flesher, " Surface Magnetization of Feromagnetic Ni(110): A Polarized Low-Energy Electron Diffraction Experiment", Phys. Rev. Lett. **43**, 728 (1979).

[40] M. Mankos, D. Adler, L. Veneklasen and E. Munro, "Electron optics for low energy electron microscopy", Physics Procedia **1** (1), 485 (2008).

[41] E. Bauer, in *Science of Microscopy*, edited by P. W. Hawkes and J. C. H. Spence (Springer New York, New York, NY, 2007), pp. 605.

[42] Rigaku Oxford Diffraction, (2019), CrysAlisPro Software system, version 1.171.41.27a, Rigaku Corporation, Oxford, UK.

[43] *SCALE3 ABSPACK*; A Rigaku Oxford Diffraction program (1.0.11,gui:1.0.7) (c) 2005-2019 Rigaku Oxford Diffraction.

[44] R. C. Clark and J. C. Reid, "The analytical calculation of absorption in multifaceted crystals", Acta Cryst. **A5**1, 887 (1995).

[45] G. M. Sheldrick, "SHELXT - Integrated space-group and crystal-structure determination", Acta Cryst. **A71**, 3 (2015).

[46] G. M. Sheldrick, Crystal structure refinement with SHELXL", Acta Cryst. **C71**, 3 (2015).

[47] O. V. Dolomanov, L. J. Bourhis, R. J. Gildea, J. A. K. Howard, and H. Puschmann, "OLEX2: a complete structure solution, refinement and analysis program", J. Appl. Cryst. **42**, 339 (2009).

[48] D. S. Kim, J. Y. Kee, J.-E. Lee, Y. Liu, Y. Kim, N. Kim, C. Hwang, W. Kim, C. Petrovic, D. R. Lee, C. Jang, H. Ryu and J. W. Choi,"Surface oxidation in a van der Waals ferromagnet Fe3-xGeTe2", Current Applied Physics **30**, 40 (2021).

[49] R. Fujita, P. Bassirian, Z. Li, Y. Guo, M. A. Mawass, F. Kronast, G. van der Laan and T. Hesjedal, "Layer-Dependent Magnetic Domains in Atomically Thin Fe(5)GeTe(2)", ACS Nano **16** (7), 10545 (2022).

[50] N. León-Brito, E. D. Bauer, F. Ronning, J. D. Thompson and R. Movshovich, "Magnetic microstructure and magnetic properties of uniaxial itinerant ferromagnet Fe3GeTe2", Journal of Applied Physics **120** (8), 083903 (2016).

[51] G. D. Nguyen, J. Lee, T. Berlijn, Q. Zou, S. M. Hus, J. Park, Z. Gai, C. Lee and A.-P. Li, "Visualization and manipulation of magnetic domains in the quasi-two-dimensional material Fe3GeTe2", Physical Review B **97** (1), 014425 (2018).

[52] M. J. Meijer, J. Lucassen, R. A. Duine, H. J. M. Swagten, B. Koopmans, R. Lavrijsen and M. H. D. Guimaraes, "Chiral Spin Spirals at the Surface of the van der Waals Ferromagnet Fe(3)GeTe(2)", Nano Lett **20** (12), 8563 (2020).

[53] B. Ding, Z. Li, G. Xu, H. Li, Z. Hou, E. Liu, X. Xi, F. Xu, Y. Yao and W. Wang, "Observation of Magnetic Skyrmion Bubbles in a van der Waals Ferromagnet Fe(3)GeTe(2)", Nano Lett **20** (2), 868 (2020).

[54] S. L. Halgedahl, "Domain Pattern Observation in Rock Magnesism: Progess and Problems", Physics of the Earth and Planetary Interiors **46**, 127 (1987).

[55] R. Szymczak, "The Magnetic Structure of Ferromagnetic Materials of Uniaxial Symmetry", Electron Technoogy **1**, 5 (1968).

[56] D. J. Craik (Editor), *Magnetic Oxides, Part 2*, (Wiley, New York, 1975), p. 483.

[57] (a) M. Mankos, D. Adler, L. Veneklasen and E. Munro,"Electron optics for low energy electron microscopy", Physics Procedia **1** (1), 485 (2008). (b)R. J. Phaneuf and A. K. Schmid,"Low-Energy Electron Microscopy: Imaging Surface Dynamics", Physics Today **56** (3), 50-55 (2003).

[58] E. Bauer, in *Springer Handbook of Microscopy* (2019), pp. 487-535.

[59] E. Bauer, in *Science of Microscopy*, edited by P. W. Hawkes and J. C. H. Spence (Springer New York, New York, NY, 2007), pp. 605-656.

[60] E. Bauer, "Low Energy Electron Microscopy", Rep. Prog. Phys. **57**, 895 (1994)

[61] L. Peng, F. S. Yasin, T. E. Park, S. J. Kim, X. Zhang, T. Nagai, K. Kimoto, S. Woo and X. Yu, "Tunable Néel–Bloch Magnetic Twists in Fe3GeTe2 with van der Waals Structure", Advanced Functional Materials **31** (37), 2103583 (2021).

[62] B. Ding, Z. Li, G. Xu, H. Li, Z. Hou, E. Liu, X. Xi, F. Xu, Y. Yao and W. Wang, "Observation of Magnetic Skyrmion Bubbles in a van der Waals Ferromagnet Fe(3)GeTe(2)", Nano Lett **20** (2), 868 (2020).

[63] G. D. Nguyen, J. Lee, T. Berlijn, Q. Zou, S. M. Hus, J. Park, Z. Gai, C. Lee and A.-P. Li, "Visualization and manipulation of magnetic domains in the quasi-two-dimensional material Fe3GeTe2", Physical Review B **97** (1), 014425 (2018).

[64] U. K. Rößler, A. N. Bogdanov and C. Pfleiderer, "Spontaneous skyrmion ground states in magnetic metals", Nature **442** (7104), 797 (2006); A. Bogdanov and A. Hubert, "Thermodynamically stable vortex states in magnetic crystals", J. Mag. Magnetic Mat. 139, 255 (1994); A. N. Bogdanov and A. Yablonskii, "Thermodynamically stable "vortices" in magnetically ordered crystals. The mixed state of magnets" JETP Lett., **68**, 101 (1989).

[65] O. Gunnarsson, M. Calanra, J. E. Han, "Saturation of Electrical Resistivity", Rev. Mod. Phys. 75, 1085 (2003).

[66] K. Kim, J. Seo, E. Lee, K. T. Ko, B. S. Kim, B. G. Jang, J. M. Ok, J. Lee, Y. J. Jo, W. Kang, J. H. Shim, C. Kim, H. W. Yeom, B. Il Min, B. J. Yang and J. S. Kim, "Large anomalous Hall current induced by topological nodal lines in a ferromagnetic van der Waals semimetal", Nat Mater **17** (9), 794 (2018).

[67] O. Y. Ridzel, V. Astašauskas and W. S. M. Werner, "Low energy (1–100 eV) electron inelastic mean free path (IMFP) values determined from analysis of secondary electron yields (SEY) in the incident energy range of 0.1–10 keV", Journal of Electron Spectroscopy and Related Phenomena **241**, 146824 (2020).

[68] H. T. Nguyen-Truong, "Electron inelastic mean free path at energies below 100 eV", J Phys Condens Matter **29** (21), 215501 (2017).

[69] Q. Li, M. Yang, C. Gong, R. V. Chopdekar, A. T. N'Diaye, J. Turner, G. Chen, A. Scholl, P. Shafer, E. Arenholz, A. K. Schmid, S. Wang, K. Liu, N. Gao, A. S. Admasu, S. W. Cheong, C. Hwang, J. Li, F. Wang, X. Zhang and Z. Qiu, "Patterning-Induced Ferromagnetism of Fe(3)GeTe(2) van der Waals Materials beyond Room Temperature", Nano Lett **18** (9), 5974 (2018).

[70] H.-H. Yang, N. Bansal, P. Rüßmann, M. Hoffmann, L. Zhang, D. Go, Q. Li, A.-A. Haghighirad, K. Sen, S. Blügel, M. Le Tacon, Y. Mokrousov and W. Wulfhekel, "Magnetic domain walls of the van der Waals material Fe3GeTe2", 2D Materials **9** (2), 025022 (2022).

# Surface Magnetism in Fe$_3$GeTe$_2$ Crystals

T. A. Tyson[1,5,*], S. Amarasinghe[1], AM M. Abeykoon[2], R. Lalancette[3], S. K. Du[4,5], X. Fang[4,5], S.-W. Cheong[4,5], A. Al-Mahboob[6], and J. T. Sadowski[6,*]

[1]Department of Physics, New Jersey Institute of Technology, Newark, NJ  07102
[2]National Synchrotron Light Source II, Brookhaven National Laboratory, Upton, NY 11973
[3]Department of Chemistry, Rutgers University, Newark, New Jersey 07102
[4]Department of Physics and Astronomy, Rutgers University, Piscataway, NJ 08854
[5]Rutgers Center for Emergent Materials, Rutgers University, Piscataway, NJ 08854
[6]Center for Functional Nanomaterials, Brookhaven National Lab, Upton, New York 11973

# (Supplementary Information Document)

*Corresponding Authors:

T. A Tyson, e-mail: tyson@njit.edu

J. T. Sadowski, e-mail: sadowski@bnl.gov

**Sample Preparation and Conditions**

Single crystals of $Fe_3GeTe_2$ were prepared by chemical vapor transport. A crystal was coated in oil, and cracked. A small crystal (220 μm x 130 μm x 60 μm) from the main crystal was cleaved for use in single crystal diffraction measurements. For the LEEM and PEEM experiments, samples were cleaved in air by the scotch tape method and transferred to the vacuum system immediately (~1 minute) after cleaving. The antechamber for sample transfer was at a pressure of $\approx 2 \times 10^{-9}$ Torr and the pressure of the UHV main chamber used in the measurements was $\approx 5 \times 10^{-10}$ Torr.

**Low Energy Electron Microscopy (LEEM) and X-ray Photoelectron Emission Microscopy Measurements (XPEEM)**

X-ray absorption spectroscopy (XAS) and low-energy electron microscopy (LEEM) measurements have been performed at the XPEEM/LEEM end station of the Electron Spectro-Microscopy beamline (ESM, 21-ID) at the National Synchrotron Light Source II. All measurements were conducted using the aberration-corrected ELMITEC LEEM III systemequipped with an R200 energy anayzer and the TVIPS TemCam XF-416UHV fiber-coupled CCD detector. X-ray absorption spectroscopy (XAS) measurements were performed in a partial electron yield (PEY) mode collecting the secondary electrons with the 2 eV energy analyzer slit centered over the maximum of the secondary electron emission peak. Pixel-wise XAS were obtained by recording a series of XPEEM images at each photon energy in a given absorption edge range at sequential increments of 0.2 eV. The LEEM measurements were conducted in mirror electron microscopy mode with the incident electrons having low kinetic energy ($\approx 1$ eV) on striking the surface of the sample. In tilted-beam mode for the LEEM measurements, the electrons strike the sample on an off-normal angle, as illustrated in Fig. 3.

**Single Crystal Diffraction Measurements**

Single crystal data sets were collected at 275 K and 100 K on a 220 μm x 130 μm x 60 μm single crystal using a Rigaku XtaLAB Synergy-S X-ray diffractometer equipped with a HyPix-6000HE hybrid photon counting (HPC) detector and microfocused Mo-Kα radiation ($\lambda = 0.71073$ Å). The HPC detector has a high dynamic range with maximum count rate = $10^6$ cps/pixel (counter depth = 31 bit, 100 μm x 100 μm pixels, Si pixels each with CMOS-based readout electronics, direct photon counting, zero dark or readout noise). Use of Mo-Kα radiation resulted in sampling depth of approximately 40 microns (full mass density). This should be compared to a sample dept of 6 microns (full density) for Cu-Kα radiation ($\lambda = 1.54184$ Å). The acquisition of the sixfold redundant datasets was conducted using CrysAlis[Pro] [1]. Subsequent data processing was also performed in CrysAlis[Pro]. Scaling (SCALE3 ABSPACK scaling algorithm [2]) and analytical absorption corrections [3] were applied to the data (based on face-indexing in CrysAlis[Pro]). The structures were solved *via* intrinsic phasing methods using ShelXT and refined using ShelXL in the Olex2 graphical user interface.

**Powder Diffraction Data**

Powder diffraction data were collected using Beamline 28-ID-1 (PDF) of the NSLS-II, with a wavelength of 0.1665 Å in a 1 mm Kapton capillary sealed with clay. We note that at this wavelength, the attention of the beam by the sample is 0.7 mm for 50% packing on the 500 mesh powder in the capillary (https://calc.weingos.com/). This bulk measurement can be compared with laboratory powder measurements using Cu Kα and Mo Kα radiation. These latter measurements only probe the surface of the samples in the holders, with penetration depths near 6 μm and 40 μm, respectively (https://henke.lbl.gov/optical_constants/atten2.html). Note that Mo Kα radiation penetration is compatible with the single crystal used in the measurement above.

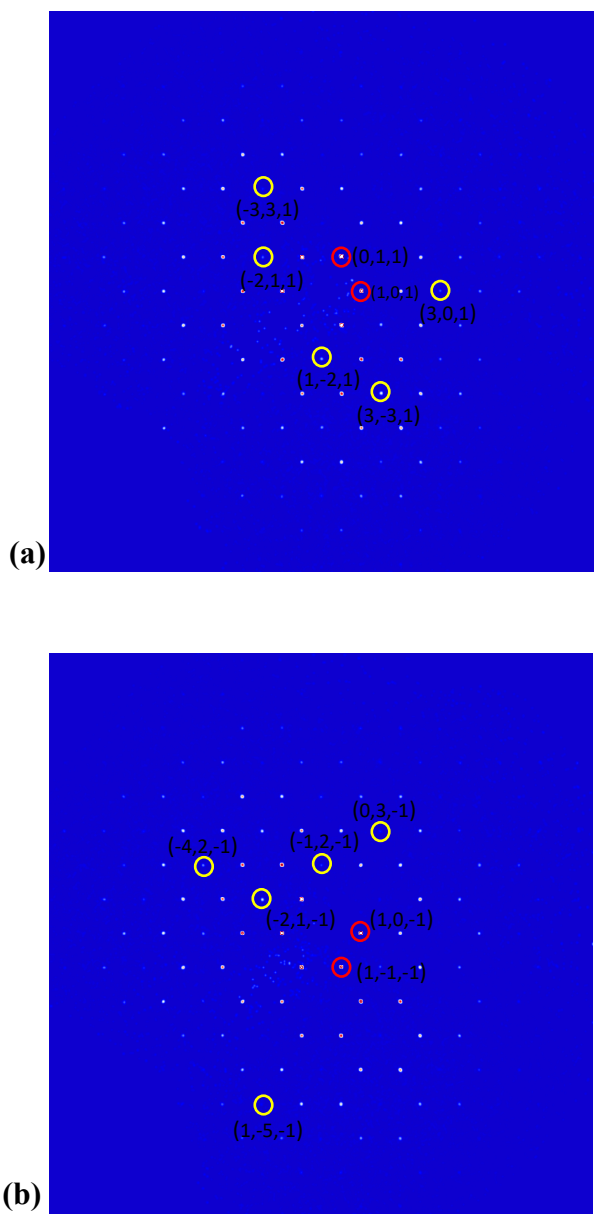

**Fig. S1.** Reciprocal space images showing (a) the ($h,k,1$) and (b) ($h,k,-1$) planes from 100 K data. Forbidden reflections relative to the P6$_3$/mmc space group are circled in yellow with the allowed ones in red. Note that allowed reflections for P6$_3$/mmc satisfy the conditions: ($h,h,\ell$), $\ell = 2n$; ($0,0,\ell$), $\ell = 2n$; ($h,k,\ell$), $\ell = 2n$, $h - k = 3n+1$ or $h - k = 3n+2$. The real space group has lower symmetry than P6$_3$/mmc. Relative to the allowed peak (0,1,1) the intensity of these forbidden peaks is at the ~5% level. These weak peaks will be difficult to detect in powder patterns unless long-time scans are collected.

**Table S1. Structural Parameters from Single Crystal Solution for $Fe_{2.87}GeTe_2$ at 100 K***

| Atoms | x(×10⁴) | y(×10⁴) | z(×10⁴) | Ueq (Å²)×10³ | Occ* |
|---|---|---|---|---|---|
| Te1 | 3333.33 | 6666.67 | 1598.6(9) | 9.7(3) | 1.0 |
| Te2 | 6666.67 | 3333.33 | 3399.3(8) | 9.9(3) | 1.0 |
| Ge1 | 6666.67 | 3333.33 | 0 | 19.9 (10) | 1.0 |
| Ge2 | 3333.33 | 6666.67 | 5000 | 25.3(11) | 1.0 |
| Fe1 | 0 | 0 | 776(2) | 9.3(5) | 1.0 |
| Fe2 | 3333.33 | 6666.67 | 0 | 8.3(11) | 0.94(3) |
| Fe3 | 0 | 0 | 5781.5(19) | 9.9(5) | 1.0 |
| Fe4 | 6666.67 | 3333.33 | 5000 | 20(2) | 0.80(3) |

| Atom | $U_{11}$(Å²)×10³ | $U_{22}$(Å²)×10³ | $U_{33}$(Å²)×10³ | $U_{12}$(Å²)×10³ |
|---|---|---|---|---|
| Te1 | 8.0(3) | 8.0(3) | 13.0(5) | 4.00(15) |
| Te2 | 9.1(3) | 9.1(3) | 11.6(5) | 4.54 (16) |
| Ge1 | 24.4(14) | 24.4(14) | 10.7(14) | 12.2(7) |
| Ge2 | 28.3(15) | 28.3(15) | 19.4(19) | 14.1(8) |
| fFe1 | 4.4(6) | 4.4(6) | 19.3(13) | 2.2(3) |
| Fe2 | 0.5(10) | 0.5(10) | 24(3) | 0.3(5) |
| Fe3 | 11.5(7) | 11.5(7) | 6.9(10) | 5.7(4) |
| Fe4 | 29(3) | 29(3) | 3(2) | 14.5(14) |

Space Group: P-6m2 (Z=2)
a = 3.98553(6) (8)Å, c = 16.2927(3) Å, Dx = 7.228 g/cm³
V = 224.128(8) Å³
Measurement Temperature: 101.1 K
Crystal Dimensions: 220 μm x 130 μm x 60 μm
Wavelength: Mo Kα (λ = 0.71073 Å)
Absorption Coefficient: 28.286 mm⁻¹
F(000) = 421.0
2θ Range : 5.0° to 80.444°
-7≤ h ≤7, -7≤ k ≤ 7,  and -29 ≤ l ≤29
Number of Measured Reflections: 40,357 (6-fold redundant over complete sphere)
Flack parameter: 0.6(4)
Independent reflections/restraints/parameters: 641/0/25
Max and Min Peak in Final Difference Map: 4.5(Te1)/-4.1(Ge2) e-/ Å³
$R_1$ = 4.64 %, $wR_2$ = 11.0 %, Goodness of Fit = 1.17(I>=2σ (I))
$R_1$ = 4.94 %, $wR_2$ = 11.3 % (all data)
$R_{int}$ = 15.5 %, $R_{sigma}$= 2.37 %

$$R_1 = \sum ||F_o| - |F_c||/ \sum |F_o|$$
$$wR_2 = \sum w(F_o^2 - F_c^2)^2)/ \sum w(F_o^2)^2$$

*Fe2 and Fe4 sites have position occupancy less than 1. All other sites have unit occupancy. Within the experimental error, the site occupancies of the 100 K and 275 K data sets are the same as expected.

**Table S2. Structural Parameters from Single Crystal Solution for Fe$_{2.85}$GeTe$_2$ at 275 K**

| Atoms | x(×10$^4$) | y(×10$^4$) | z(×10$^4$) | Ueq (Å$^2$)×10$^3$ | Occ* |
|---|---|---|---|---|---|
| Te1 | 3333.33 | 6666.67 | 1597.1(7) | 15.5(3) | 1.0 |
| Te2 | 6666.67 | 3333.33 | 3400.0(7) | 15.8(3) | 1.0 |
| Ge1 | 6666.67 | 3333.33 | 0 | 24.7(10) | 1.0 |
| Ge2 | 3333.33 | 6666.67 | 5000 | 29.0(11) | 1.0 |
| Fe1 | 0 | 0 | 778.3(17) | 14.2(4) | 1.0 |
| Fe2 | 3333.33 | 6666.67 | 0 | 11.2(10) | 0.88(2) |
| Fe3 | 0 | 0 | 5783.2(16) | 15.2(4) | 1.0 |
| Fe4 | 6666.67 | 3333.33 | 5000 | 21.6(16) | 0.83(3) |

| Atom | U$_{11}$(Å$^2$)×10$^3$ | U$_{22}$(Å$^2$)×10$^3$ | U$_{33}$(Å$^2$)×10$^3$ | U$_{12}$(Å$^2$)×10$^3$ |
|---|---|---|---|---|
| Te1 | 14.6(3) | 14.6(3) | 17.2(4) | 7.32(15) |
| Te2 | 15.7(3) | 15.7(3) | 16.0(4) | 7.84(16) |
| Ge1 | 27.0(13) | 27.0(13) | 20.1(17) | 13.5(6) |
| Ge2 | 30.0(13) | 30.0(13) | 27(2) | 15.0(6) |
| Fe1 | 10.1(5) | 10.1(6) | 22.5(11) | 5.0(3) |
| Fe2 | 5.7(9) | 5.7(9) | 22(2) | 2.8(5) |
| Fe3 | 16.0(7) | 16.0(7) | 13.6(9) | 8.0(3) |
| Fe4 | 27(2) | 27(2) | 11(2) | 13.5(11) |

Space Group: P-6m2 (Z=2)
a = 3.9933(2)Å, c = 16.3461(2) Å, Dx = 7.160 g/cm$^3$
V = 225.740(3) Å$^3$
Measurement Temperature: 275 K
Crystal Dimensions: 220 μm x 130 μm x 60 μm
Wavelength: Mo Kα (λ = 0.71073 Å)
Absorption Coefficient: 28.023 mm$^{-1}$
F(000) = 420.0
2θ Range : 7.478° to 80.458°
-7≤ h ≤7, -7≤ k ≤ 7, and -29 ≤ l ≤29
Number of Measured Reflections: 41,173 (6-fold redundant over complete sphere)
Flack parameter: 0.3(4)
Independent reflections/restraints/parameters : 644/0/25
Max and Min Peak in Final Difference Map: 3.8(Ge1)/-3.3(Ge1) e-/ Å$^3$
R$_1$ = 4.31 %, wR$_2$ = 11.6 %, Goodness of Fit = 1.15(I>=2σ (I))
R$_1$ = 4.54 %, wR$_2$ = 11.8 % (all data)
R$_{int}$ = 14.1%, R$_{sigma}$= 2.20 %

$$R_1 = \sum ||F_o| - |F_c||/\sum |F_o|$$
$$wR_2 = \sum w(F_o^2 - F_c^2)^2)/\sum w(F_o^2)^2$$

*Fe2 and Fe4 sites have position occupancy less than 1. All other sites have unit occupancy.

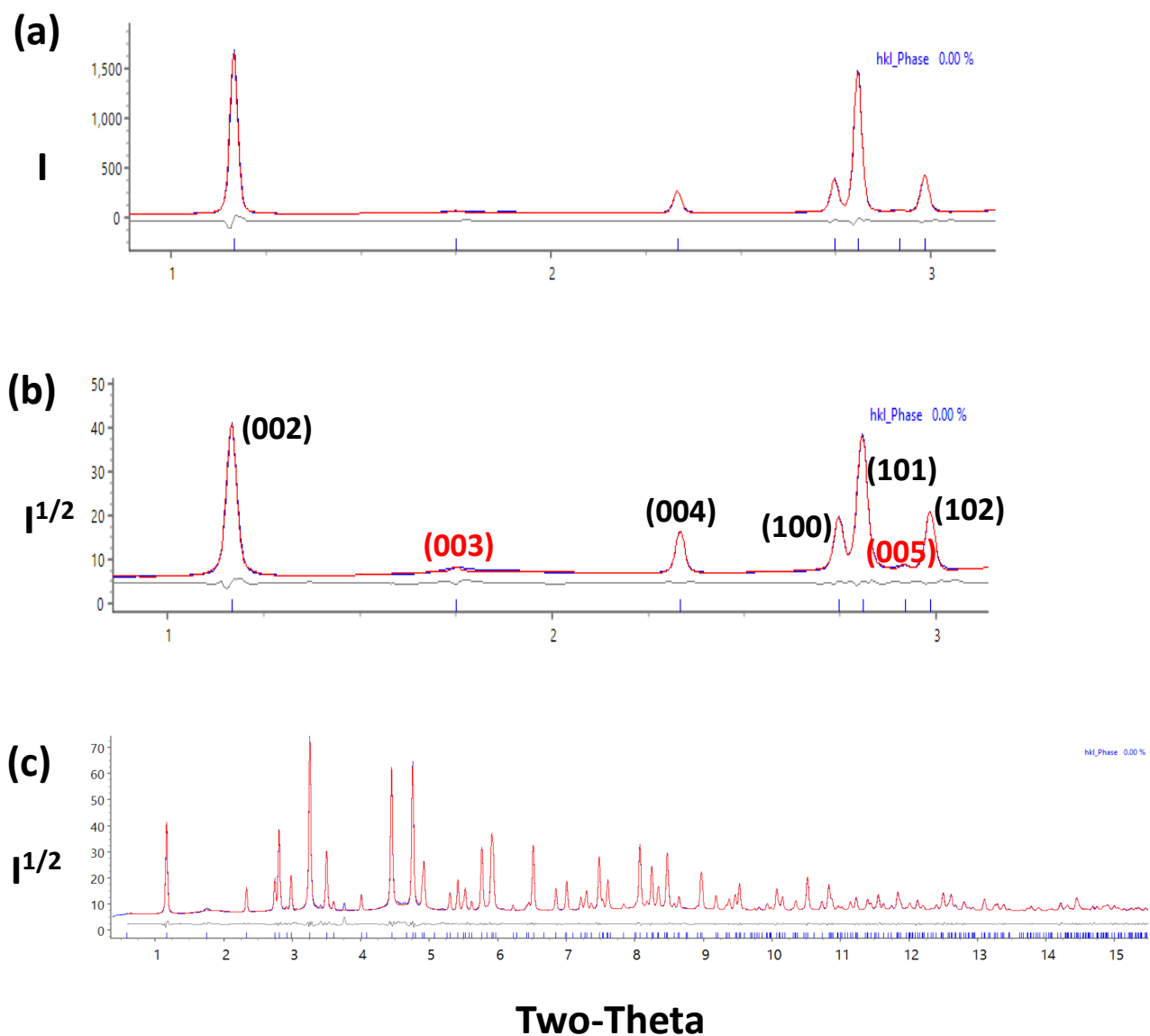

**Fig. S2.** Low-angle synchrotron x-ray powder diffraction patterns (at 304 K) measured with wavelength $\lambda$ = 0.1665 Å (in panels (a) and (b)). Panel (a) shows the intensity vs. two-theta plot, while to access the weak peaks we show the square root of the intensity vs two-theta in (b). The residual curve is presented as the grey curve, while the data and LeBail fits for the P-6m2 space group are presented as the blue and red curves, respectively. In (b) the peaks with $(h,k,l)$ indices in red are forbidden in the $P6_3/mmc$ spacegroup. The full data set with fit is shown in panel (c).

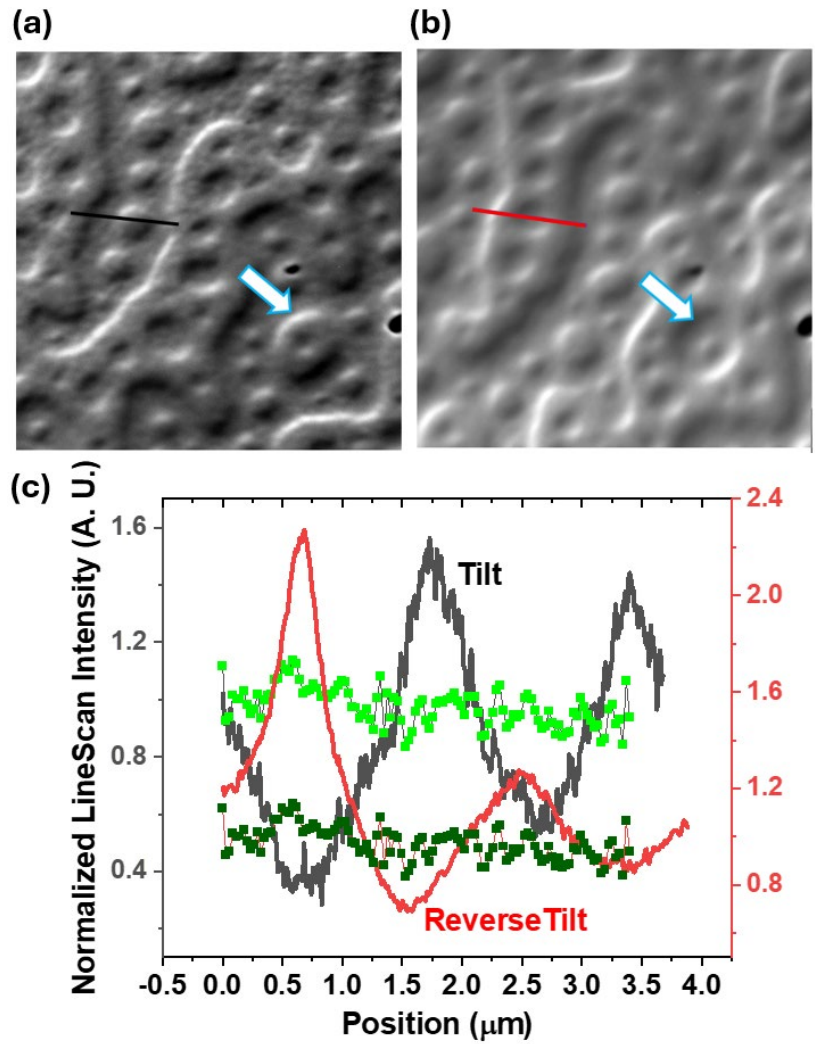

**Fig. S3.** Tilted-beam LEEM measurements taken from 13 μm x 13 μm area of the sample with (a) tilt and (b) reverse tilt configurations showing the reversal of intensity in the domain wall regions as predicted in Fig. 3. The arrow points to the same domain in both images; (c) Line scans showing the relative intensities of lines in the domains for both beam directions. The additional green curves are for the untilted beam on the y-axis scale of the tilted (light green) and reverse tilted (dark green) configurations.

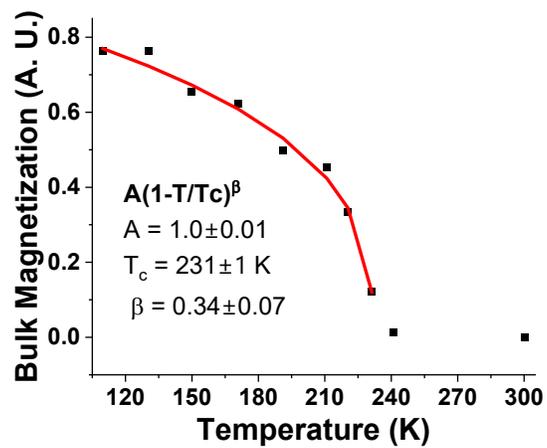

**Fig. S4.** Power law fit $(A*(1-T/Tc)^\beta)$ to the bulk magnetization data (black points taken from SQUID measurements of Li *et al*. (Nano Lett. 18, 5974 (2018) with background subtracted). For this bulk data, the fit was conducted between 170 K and 231 K and extrapolated to lower temperatures (red line).

# References

[1] Rigaku Oxford Diffraction, (2019), CrysAlisPro Software system, version 1.171.41.27a, Rigaku Corporation, Oxford, UK.

[2] *SCALE3 ABSPACK*; A Rigaku Oxford Diffraction program (1.0.11,gui:1.0.7) (c) 2005-2019 Rigaku Oxford Diffraction.

[3] R. C. Clark and J. C. Reid, Acta Cryst. A51, 887 (1995).